\documentclass[12pt]{article}
\usepackage{subfigure}
\usepackage{epsfig}

\input epsf
\def\esp{\mathrm{e}}
\def\beq{\begin{eqnarray}}
\def\eeq{\end{eqnarray}}

\def\L*{{\cal L}_*}
\def\lsim{\mathrel{\rlap{\lower3pt\hbox{\hskip0pt$\sim$}}
     \raise1pt\hbox{$<$}}}         
\def\gsim{\mathrel{\rlap{\lower4pt\hbox{\hskip1pt$\sim$}}
     \raise1pt\hbox{$>$}}}         

\usepackage{amsmath}
\usepackage{amsfonts}
\usepackage{amssymb}

\begin{document}

\begin{titlepage}

\thispagestyle{empty}

\begin{flushright}
{NYU-TH-08/06/11}
\end{flushright}
\vskip 0.9cm

\centerline{\Large \bf  Charged Condensate and Helium 
Dwarf Stars}                    

\vskip 0.7cm
\centerline{\large Gregory Gabadadze$^{a,b}$ and  Rachel A. Rosen$^{a}$}
\vskip 0.3cm
\centerline{\em $^{a}$Center for Cosmology and Particle Physics,  
Department of Physics}
\centerline{\em New York University, New York, 
NY  10003, USA}
\centerline{\em $^{b}$Joseph Henry Laboratories, Princeton University, 
Princeton, NJ  08544, USA}

\vskip 1.9cm

\begin{abstract}
White dwarf stars composed of carbon, oxygen or heavier elements 
are expected to crystallize as they cool down below certain temperatures.   
Yet,  simple arguments suggest that 
the helium white dwarf cores may not solidify, mostly because of 
zero-point oscillations of the helium ions that  would 
dissolve the crystalline structure.  We argue  that the interior 
of the helium  dwarfs may instead form a macroscopic quantum state 
in which the  charged helium-4 nuclei are in a Bose-Einstein condensate, 
while the relativistic electrons form a neutralizing degenerate Fermi 
liquid.  We discuss the electric charge screening,  and the 
spectrum of this substance, showing that the  bosonic long-wavelength  
fluctuations  exhibit a mass gap.  Hence, there is a suppression 
at low temperatures of the boson contribution to the specific heat --
the latter being  dominated by  the specific heat of the 
electrons near the Fermi surface.  This state of matter 
may have observational signatures.

\end{abstract}

\vspace{0.5in}

\end{titlepage}

\newpage

\section {Introduction}

Consider a system of a large number of neutral atoms, each having  
$Z$ electrons, and for  simplicity we focus on  $Z\leq 10$. 
Suppose  we increase the number-density of 
the atoms -- denoted by $J_0/Z$, where $J_0$ is the electron 
number-density -- so that  the average separation 
between the atomic nuclei
\beq
d \equiv \left ( 3 Z \over 4 \pi   J_0 \right )^{1/3}\,,
\label{d}
\eeq
becomes much smaller than the Bohr radius,  
but is still much greater than the size of the  charged  nuclei.
This would certainly be the case when, for instance, 
$J_0\simeq (0.5-5~MeV)^3$, for which  the electron  
Fermi momentum,  $p_F = (3 \pi^2 J_0)^{1/3}$, ranges  
from well-relativistic to  ultra-relativistic.

Under these circumstances the electron Fermi energy $E_F$ 
will exceed the atomic binding energy, and 
the neutral atoms will dissolve into the electrons and nuclei.
What is the ground state of such a system, 
when it's held together by gravity?

The answer would depend on its temperature $T$.  At certain high temperatures 
a plasma of negatively  charged electrons and  positively charged 
nuclei (ions) would be the equilibrium state. However, for lower temperatures
the system would undergo significant charges. 

Let us consider the cooling process by ignoring the 
quantum mechanical effects for the nuclei (ions), while  treating the 
electrons as a degenerate Fermi liquid\footnote{The Coulomb 
energy of a pair of electrons here is smaller than  the Fermi energy.}.  
Consider  the  nuclei that are bosons. The strength of interactions of a 
{\it classical} plasma of the charged nuclei (ions) is customarily  
characterized  by  the ratio of the average Coulomb energy of a pair 
of ions to the thermal energy (see, e.g., \cite {Shapiro}) 
\beq
\Gamma\equiv {E_{Coulomb}\over 2E_{Thermal}/3}= {(Ze)^2/(4 \pi d)\over k_B T},
\label{Gamma}
\eeq
here $e$ denotes the electric charge, 
$k_B$ is the Boltzmann constant, and the second equality in 
(\ref {Gamma}) assumes the validity of the classical 
approximation.

It has been known that  when  $\Gamma \gsim 180$  
the ion plasma is coupled  strongly-enough for 
the system to  crystallize, i.e.,  when the   
temperatures become low enough and $\Gamma$ reaches 
the above-mentioned value, the system would undergo 
crystallization transition (for earlier works, see Refs. 
\cite {Ruderman,vanHorn}, for latter studies,  
\cite {Ichi,DeWitt,Chabrier} and references therein).  
This has direct relevance to white dwarf stars (WDs): 
It is expected that in most of 
the white dwarfs, consisting of carbon, oxygen, or heavier elements, 
the crystallization transition takes place in the process of 
cooling (for a review, see, e.g., Ref. \cite {WDs}). 

The above discussions were classical.
Although it was observed in  Ref.  \cite {Ashcroft} that the quantum 
effects of the {\it ions} may be important for cooling of the WDs, 
quantitative studies using a model for a quantum solid and   
Lindemann's empirical rule  lead to a 
conclusion  \cite {Chabrier} that the  melting curves for most of the WDs 
are not changed significantly by the  quantum effects.  

\vspace{0.1in}

In the present work we're  interested in the system of 
electrons and helium-4 nuclei. The latter being lighter than the other 
nuclei,  and having lower charge, should be expected 
to reveal stronger sensitivity to the quantum effects. 

Such a system should be present in cores of  
helium white dwarfs. These dwarf stars stayed  
relatively cool due to their binary companions,  
which kept  accreting  a part of the dwarf's energy 
during their paired evolution.  As a result, the temperature in 
such  dwarfs was never enough to begin the helium burning  process. 

The electron number-densities in the interval $J_0 \simeq (0.5-5~MeV)^3$
would correspond to the mass density of the 
helium-electron system $5\cdot (10^7 - 10^{10})~g/cm^3$.
The densities closer to the  lower end  of this 
interval (which we refer to as  the low densities hereafter)
may well exist in cores of the helium white dwarfs.
We will also discuss the densities closer to the 
upper end, as they may be present in some other astrophysical 
circumstances.  

The system should be long lived as the helium-4  nuclei are 
stable w.r.t. fission. Furthermore, some nuclear reactions that 
could contaminate the helium-4 cores by their products are 
suppressed.  One of this is the neutronization process 
due to the inverse beta-decay. In our case 
the electrons with  $J_0\simeq (0.5-5~MeV)^3$ are not energetic 
enough to reach the  neutronization threshold of 
the helium-4 nucleus,  which is about $20~{\rm MeV}$.
Moreover, we would expect that the rate of helium 
fusion via the triple alpha-particle reaction is 
suppressed  as the so-called 
pycnonuclear reaction (i.e., nuclear fusion due to zero-point oscillations 
in a  high density environment) rates are exponentially small \cite {Shapiro}.
Hence,  the cores of these WD's are expected to be dominated 
by  helium-4 for very long time after their formation.

In what follows, we will focus on the case when  $Z=2$.  
Then, for $J_0\simeq (0.5-5~MeV)^3$, the classical 
considerations of Eq. (\ref {Gamma})  would give 
the crystallization temperature of the helium ions 
$T_{\rm cryst} \simeq  (10^6 - 10^7)~K$.
However, before these temperatures are reached from above,  
the de Broglie  wavelengths  of the helium nuclei would start to overlap, 
suggesting that the quantum effects of the ions may be 
important.  An estimate for the critical temperature below 
which this will happen can be given by 
\beq
T_c \simeq { 4\pi^2 \over 3 m_H d^2}\,.
\label{dB}
\eeq
For the number densities at hand  we obtain  $T_c \simeq  
(10^7 - 10^9)~K$. These are greater than  
the crystallization  temperatures estimated above.  For this, 
it's unjustified to approximate the thermal energy of the ions
by its classical expression $3k_BT/2$, as it was done in 
the denominator of (\ref {Gamma})\footnote{Note that  
for the system of carbon nuclei we would obtain the crystallization 
temperature $T_{\rm cryst} \simeq  6\cdot (10^6  - 10^7)~K$, and 
the corresponding  critical temperatures $T_c \simeq  2\cdot (10^6 - 10^8)~K$. 
Thus, for the low densities the crystallization temperature  exceeds the 
critical one, and the carbon solidification would take place. For the 
higher densities  the opposite statement would be true. One should 
note, however, that  we're comparing these for a fixed electron 
number-density in the helium and carbon cases, which implies three 
times greater mass density for the carbon case, so the comparisons should 
be readjusted by this factor. Since the carbon cores is  not a 
subject  of the present paper, we will not deal with this procedure 
here.}.

Perhaps the most important quantum effect, that invalidates 
the predictions of (\ref {Gamma}), is related to the 
zero-point fluctuations: Even at very 
low temperatures, close to the absolute zero, 
there will be a significant amount of energy stored in  
the quantum mechanical  zero-point oscillations of the nuclei 
in a would-be crystal.   This energy  
can be approximated as follows (see, e.g., \cite {Shapiro}):
\beq
E_0 ={3\over 2}\omega^{ions}_0 \equiv 
{3\over 2} {\omega^{ions}_p\over \sqrt{3}} 
= \left ( { 3 J_0 (Ze)^2 \over 4 m_H Z} 
\right )^{1/2}\simeq ( 3 \cdot 10^3 -  10^5)~{\rm eV}\,,
\label{zp}
\eeq
where $m_H \simeq 3.7 ~{\rm GeV}$ is the helium nucleus mass, and 
$\omega^{ions}_p$ denotes the ion plasma frequency. 
This energy is commensurate with the  
classical thermal energy at temperatures 
$(2/3)(3 \cdot 10^7 -  10^{9})~{\rm K}$ -- the  values that 
are greater than the crystallization temperatures 
obtained from ({\ref {Gamma}). 

The above discussions suggest that  a 
measure of coupling in such a cold plasma is the ratio 
of the Coulomb energy to  the zero-point energy.
This ratio, properly normalized to be consistent with 
the conventions of \cite {Shapiro} (which are different from ours), 
takes a small value that is not  enough for the plasma to 
crystallize. It increases  with the nuclear charge $Z$,  and the atomic  
number $A$.  Thus, for heavier nuclei, it will  
be high-enough not to obstruct  the crystallization. 
In the charged helium-4 case with the densities at hand, 
however, the zero-point oscillations would melt the crystal 
lattice\footnote{The case for the
crystallization is worsened further by the fact that for the 
relativistic electrons with the Fermi energy $E_F\simeq (3\pi^2 J_0)^{1/3}$,
the  Coulomb interaction gets screened.
The screening in such a plasma becomes effective at
scales of the order the Debye wavelength,
$\lambda_D\sim (E_F/ e^2J_0)^{1/2} \sim (e^3 J_0)^{-1/3}$, and
can be appreciable at the scale of the average
inter-particle separation $d$.  Hence, not only the denominator
in (\ref {Gamma}) should be modified,
but also the numerator should be reduced (multiplied by a factor) 
to account for the screening. The question of how significantly 
in the (ultra)relativistic case this screening may  
affect the crystallization of  the carbon nuclei is left open.  
Note that this screening  can be important in a nonrelativistic 
setup as well, see, e.g., Ref. \cite {Ichi}.}.

The above considerations lead us to conclude that the dense system of the 
helium-4 nuclei and electrons may not solidify, mostly 
because of the zero-point oscillations of the helium  nuclei.

\section {Charged condensation}

What could  then be an adequate description of the quantum state
discussed above?   It was argued in Ref. \cite {GGRR1} that 
a state in which the charged helium-4 nuclei 
form a  Bose-Einstein (BE) condensate, while the degenerate 
electrons neutralize the net charge of the condensate,
represents a local equilibrium  state of the system.

The helium nuclei could have condensed to such a state
when their  de Broglie wavelength started to overlap,
which according to our estimates above happened 
before the temperature  dropped to the would-be crystallization value. 
Such a state of the helium  nuclei cannot be resolved  
into individual  particles. It represents a macroscopic 
mode with a large occupation number, somewhat similar to 
a quantum liquid of charged particles, where the charge 
is screened.  This state, referred to as charged condensate, 
has  a number of distinctive signatures:
a photon propagating through this medium slows down, since it  
acquires a Lorentz-violating mass term,  and its longitudinal 
polarization exhibits unusual  dispersion relation. Moreover, 
electric charges in this medium, were shown in \cite {GGRR1},  
are screened at the distance scales $1/M$, where $M$ for 
the helium-electron system denotes 
\beq
M \equiv \left (2e^2 m_H J_0\right)^{1/4}\,.
\label{M} 
\eeq
Importantly, the screening takes place at scales 
larger than $1/M$;  on the other hand, $1/M < d$  
(see discussions of this in Section 3)!

In Ref. \cite {GGRR1}, however, the dynamics of the 
electrons was neglected, since the fermions considered 
there were heavier. One of the goals of the present work  
is to include the  dynamics of electrons.  We will show that 
the screening  length can be approximated by (\ref {M}), 
as long as  $m_H \gg J^{1/3}_0$ (as it is the case here).
However, we will also see that, inclusion of the 
electron dynamics modifies quantitatively the spectrum of 
small perturbations above the ground state\footnote{It was argued in Ref. 
\cite {GGRR2}, that there could  
exist long-lived  spherically symmetric  objects of a (sub)atomic 
size  with the charged condensate  of helium-4 nuclei and electrons in 
their interior that could be held together by electrodynamic forces, as 
long as there is a significant amount of uncompensated charge at the 
surface of such objects. This interesting possibility and it applications 
will be studied further elsewhere. Here, the force that keeps the 
system together is gravity.}.

\vspace{0.1in}

Before turning to these issues, let us deal with the 
gravitational part of the problem.  From now on we consider 
the  system at temperatures much lower  than  
$\sim 10^{7}~K$, and ignore thermal effects. 
Thus, we discuss  the electrons  
in the gravitational and  electrostatic fields in the interior of an 
astrophysical object, e.g., in the core of a helium dwarf. 
The Fermi momentum, $p_F$, is related to the chemical 
potential $\mu$ as follows:
\beq
\sqrt{p_F^2 +m_e^2} +  V_g + e A_0 = \mu\,,
\label{mu}
\eeq
where $m_e$ is the electron mass,  
$V_g$ is the (negative) gravitational potential 
energy per electron, and $A_0$ denotes the gauge potential.
An average  $A_0$ would be  nonzero 
for a nonzero net surface  charge.  
However, we will assume that the net charge, 
even if present, 
is small enough, so that $|eA_0| \ll |V_g|$.
Then, as is well known, the overall stability of a dwarf 
star  is due to the balance  between attractive gravity and repulsive 
Fermi degeneracy pressure of the electrons. Thus, we briefly 
discuss the interplay between $V_g$ and the Fermi energy. 

For this one could use the Thomas-Fermi (TF) approximation: 
Ignore $A_0$ in (\ref {mu}), and deduce 
from it  $p_F= ((\mu - V_g)^2-m_e^2)^{1/2}$.  
Relate the latter to the  number density 
\beq
J_0 = {p^3_F \over 3\pi^2}= {( (\mu - V_g)^2-m_e^2)^{3/2}\over 3\pi^2}\,,
\label{J0}
\eeq
and use this expression in the Poisson equation
for $V_g$  to derive 
the Chandrasekhar equation\footnote{Since this equation and  
its consequences are well studied, we'll not discuss 
them  here. For its  derivation in the TF approach, and 
further details see, Ref. \cite {Straumann}, and references therein.}.

Our main interest  is in local properties of 
particles at  scales that are much smaller than the size 
of the star. These  properties are determined by  electromagnetic 
interactions, once the average effective 
chemical potential for the particles in a given homogeneous 
region of the star is set by the gravitational dynamics.  
For instance, consider a  small region in the core of the star. 
Local distortions in the number-density (i.e., 
distortions in  the local Fermi momentum $p_F(x)$) will cause 
the respective change in the local field potential $A_0(x)$, 
while the change in the local gravitational field is 
negligible since the gravity is so 
much weaker.

These arguments can be formalized by the following relation:
\beq
\sqrt{p_F^2(x) +m_e^2} + e A_0(x) = \langle 
\mu-  V_g(x) \rangle\equiv \mu_{f} \,,
\label{mueff}
\eeq
where the brackets, $\langle ...\rangle $, denote the 
averaging of the large-scale gravity effects, over the 
region of a roughly uniform number density in the star.
Hence, by $\mu_{f}$  we denote the effective  chemical potential 
that governs the fermion dynamics in a local approximately 
homogeneous region, e.g., in the core of the star. The size of the
homogeneous region should be greater than the photon Compton 
wavelength, for our discussions below to be valid.

\vspace{0.1in}

Having the gravity part dealt with,  let us  
turn to the microscopic non-gravitational physics.
For a large number of quanta that form  a macroscopic state, such 
as the charged condensate, the description in terms of 
the effective field theory of the order parameter, and its 
long wavelength fluctuations, becomes adequate.  
The effective field theory  will be constructed based on 
the  fundamental principles of  symmetries and properties of 
the interactions involved. This is the formalism that we will adopt for 
the rest of the work.

The effective Lagrangian that captures the  microscopic  
properties of the system contains the photon  
field $A_\mu$,  a charged scalar field $\phi$,  and 
fermions $\Psi^+,\Psi$, with the effective chemical potential 
$\mu_f$ discussed above
\beq
{\cal{L}} = -{1\over 4} F_{\mu\nu}^2 + 
\vert {\tilde D}_{\mu} \phi \vert^2 - m_H^2 \phi^{\ast} \phi +
{\bar \Psi}(i\gamma^\mu D_\mu -m_e)\Psi + \mu_{f} \Psi^+ \Psi\,.
\label{lagr0}
\eeq
In a conventional setup, upon quantization,  the fluctuations of 
$\phi$ would describe the helium-4  nucleus of charge $+2e$,  
and its antiparticle. Such a description  would be 
valid at densities that are above the atomic, but well below 
the nuclear scales.  Here, instead, the vacuum expectation value of 
the field $\phi$ will serve as an order parameter for the 
condensation of the helium-4 nuclei, thus describing a 
state with a large occupation number. Because of this large number, 
the description of such a quantum state in terms of the  
classical field is valid. Fluctuations of the order 
parameter will be discussed below.

The fermions in (\ref {lagr0}) describe the  (quasi) electrons with 
charge $-e$, and (quasi) positrons.  The covariant derivative for them 
is defined  as $D_\mu = \partial_\mu +ie A_\mu $, while for the 
scalar it reads
\beq
{\tilde D}_\mu \equiv  \partial_\mu -2ie{\tilde A}_\mu \equiv
\partial_\mu -i(2eA_\mu +\mu_s \delta_{\mu0})\,.
\label{tildeD}
\eeq
The latter is equivalent to the introduction of the  chemical potential 
$\mu_s$ for the helium-4 nuclei.

The Lagrangian (\ref {lagr0}) has a global $U_s(1)$ symmetry,
yielding the conservation of the number of scalars,  
as well as another global  $U_f(1)$, responsible for  the fermion 
number conservation. One linear combination of these 
two  symmetries is gauged, and the corresponding conserved 
current is coupled to the photon 
field. We could add the quartic scalar self-interaction 
term to (\ref {lagr0}), but this won't change our results 
significantly,  as long as the quartic coupling is not strong, and 
$m_H\gg J_0^{1/3}$.

Since the electrons are relativistic, the quantum loops would 
renormalize the effective Lagrangian (\ref {lagr0}) by generating additional 
terms proportional to the fermion chemical potential and  temperature, 
which  would be  still consistent with  the symmetries of the theory. 
These terms would come suppressed by  
the fine structure constant, $e^2/4\pi$, and we do 
not expect them to modify our quantitative results significantly
in the regime when $m_H\gg J_0^{1/3}$. Hence, for simplicity of presentation 
we do  not include those terms here.

For further convenience, we introduce the following notation for 
the scalar field, $\phi = \tfrac{1}{\sqrt{2}} \sigma\, 
e^{i \alpha}$, and  work in 
the unitary gauge where the phase of the scalar is set to zero,
$\alpha=0$. In this gauge, the Lagrangian density reads: 
\beq
{\cal{L}}=- {1 \over 4}F_{\mu\nu}^2 + 
{1 \over 2}(\partial_{\mu}\sigma)^2+
{(2e)^2\over 2} {\tilde A}_\mu^2 \sigma^2- {1 \over 2}m_H^2 
\sigma^2 + {\bar \Psi}(i\gamma^\mu D_\mu -m_e)\Psi +\mu_f 
{\Psi^+} \Psi \,.
\label{lagr}
\eeq
The kinetic term for the scalar in (\ref{lagr0}) 
gives rise to the third term in the Lagrangian (\ref{lagr}).  
If the field ${\tilde A}_0$ acquires an expectation value, 
this term plays the role of a tachyonic mass for the scalars
(see, e.g., \cite {Linde}).  In particular, when 
$\langle  2e {\tilde A}_0 \rangle = m_H$, 
the scalar field condenses. That is to say, the equations of motion 
that follow from (\ref {lagr}) have the following static solution:
\beq
\langle 2e A_0 \rangle +\mu_s= m_H \,,~~~~~~~
\langle \sigma \rangle  = \sqrt{\frac{J_0}{2 m_H}} \,.
\label{b0prime}
\eeq
What fixes the value of $A_0$ is the net surface charge.  
In particular, when the surface charge is absent, 
$\langle 2e A_0 \rangle = 0$, and has to be 
$\mu_s = m_H$, for the condensation to occur.
The system is neutral in the bulk, as the charge in the condensate 
is exactly canceled by the electron charge \cite {GGRR1}.

Dynamically, the condensation proceeds in a conventional manner:
At temperatures higher than  $T_c$ most of the states are in thermal 
modes, and $\mu_s(T)$ is less than   $m_H$,  but increases as the 
temperature drops\footnote{Here we ignore the  
temperature dependence of $m_H$ which would be present due to the 
quantum loop effects.}. As $T \to T_c$ a significant fraction of the 
modes ends up in the zero momentum ground state, 
while $\mu_s(T_c)$ asymptotes to  $m_H$ (for a discussion of 
relativistic BE condensation, see, e.g., \cite {Haber}).

The last comment in this section is that we're looking at 
the  effects of the condensate that are $1/ m_H$  suppressed. 
In particular, the order parameter of the condensation that we're discussing  
(given in eq. (\ref {b0prime})) is the $1/ m_H$  suppressed quantity. 
Clearly, in  order for the condensate  to make sense
at some fixed nonzero temperature $T$, 
the de Broglie wavelength of the bosons, 
$\sim {1/\sqrt{m_H T}}$,  should be larger than the inter-particle 
separation -- the condition which would break down for 
very large $m_H$ (in agreement with the decoupling), 
but is fulfilled for the values of the 
parameters in our case. Furthermore, the inverse of  the order parameter 
$\langle \sigma \rangle$ should be smaller that the physical 
size of the system -- the condition that would also 
be  violated for some very large $m_H$ (consistent with the 
decoupling), but is preserved in our case.

\section{Charge screening and small perturbations}

As a next step we'd like to discuss screening of a probe 
electric charge in the condensate.  To determine the 
screening length we consider a  small, 
spherically symmetric object with a nonzero charge placed in the 
condensate.  Outside of the charge, in the condensate, 
the equations of motion  for a static ${A_0}$ and $\sigma$, 
as derived from (\ref{lagr0}) are:
\beq
-\nabla^2 {\tilde A}_0 + 4e^2 \sigma^2  {\tilde A}_0 = e J_0 \,,~~~~~
-\nabla^2 \sigma = (4e^2 {\tilde A}_0^2 - m_H^2) \sigma \,.
\label{eqom}
\eeq
Following \cite {GGRR1} we parametrize the perturbations as follows:
\beq
{\tilde A}_0(r) = \frac{m_H}{2e} +\delta A_0 (r) \, , ~~~~ \,  
\sigma(r) = \sqrt{\frac{J_0}{2 m_H}}+\delta \sigma(r) \,.
\eeq
To include the effects of the fermion  fluctuations 
we substitute  $J_0 =(p_F^3/3\pi^2)$ on the r.h.s. of the first equation 
in (\ref {eqom}), and express $p_F$ in terms of  the gauge potential
$A_0$, and the effective chemical potential $\mu_f$ via (\ref {mueff}). 
The result is that  $\delta J_0 (r)$
gets related to $\delta A_0$.  Thus, the coefficient in front of 
$\delta A_0$ in the final equation for the perturbations 
gets modified as compared to \cite {GGRR1}
\beq
-\nabla^2 \delta A_0 + e^2 \left [{2J_0\over m_H}  + {(3\pi^2 J_0)^{2/3}
\over \pi^2} \right ]\, 
\delta A_0 = -2 \,M^2 \,\delta \sigma \,,
\label{beqn}
\eeq
\beq
-\nabla^2 \delta \sigma = 2 \,M^2 \,\delta A_0 \,.
\label{seqn}
\eeq
The above approximations is valid as long as 
we focus on solutions that satisfy  $\delta 
\sigma \ll \sqrt{J_0/2m_H} $ and $\delta A_0 \ll m_H/2e$. 

In the regime when $m_H \gg J_0^{1/3}$, we find that 
$M^2\gg J_0^{2/3}$, and then we can neglect the 
second term on the l.h.s.  in (\ref{beqn}).  
For large $r$ we require that 
$\delta A_0, \, \delta \sigma \rightarrow 0$.  The 
solutions are similar to those of \cite {GGRR1}, in which  
the boundary conditions select the decaying functions:
\beq
\delta A_0(r) = \frac{\esp^{-Mr}}{r} 
\left[c_1 \sin(M r)+c_2 \cos(M r)\right] \,,~~~~
\label{bsol}
\eeq
\beq
\delta \sigma(r) = 
\frac{\esp^{-Mr}}{r} \left[-c_1 \cos(M r) +c_2 \sin(M r)\right] \,.
\label{ssol}
\eeq
The constants $c_1$ and $c_2$ are to be determined by matching these 
solutions to those in the interior of the small, charged object.  
Thus, for a probe particle, the screening occurs at scales greater than  
$1/M$.  For the  helium-electron system at hand, $1/M$  is larger  
than the size of the helium nuclei, but is shorter that the  
average inter-particle separation $d$. The latter statement  
needs some qualifications.  
The effective field theory description adopted above 
is not expected to hold  at scales below $d$.  
There is an explanation of this in terms 
of a cancellation between the potentials due to the   
two long-wavelength modes -- Coulomb  and ``phonon''   
quasiparticles --   both of which are much lighter than 
$1/d$,  and are well-within the validity of the effective 
field theory. This and related issues 
will be discussed in detail elsewhere.  Here  we only 
consider  phenomena at scales greater than $d$,
where screening (\ref {bsol}), (\ref {ssol})  is significant, and 
the effective field theory descriptions is valid.

The magnetic interactions for 
the fermions filling the Fermi sphere, are not screened at scale 
$M^{-1}$. Instead, as it will be clear from the Lagrangian 
(\ref {pert}) derived below,  the magnetic interactions 
are screened  at scale of the Compton wavelength of the 
massive photon.  This scale will end up being 
much  greater that the average inter-particle 
separation $d$.

\vspace{0.1in}

Let us now turn to the spectrum  of small perturbations. The electrons fill
the Fermi surface with the Fermi energy $E_F$. 
As to the bosons,  there are  three components of the massive photon,
and the heavy scalar mode.  We ``integrate out'' the fermions by  
taking into account  their effect via the TF approximation,  
like we did it above for the static case. The resulting Lagrangian 
density for the bosonic perturbations takes the form:

\beq
{\cal L}_{\rm bos} = - {1\over 4} f^2_{\mu\nu}
+ {1\over 2} m_0^2 (\delta A_0)^2 - {1\over 2} m_\gamma^2 (\delta A_j)^2 +
{1\over 2} ( \partial_\mu \delta \sigma)^2 + 2M^2 \delta  A_0  
\delta \sigma \,,
\label{pert}
\eeq
where $f_{\mu\nu}$ denotes the field strength  for 
$\delta A_\mu $, $j=1,2,3$ is the spatial index and 
\beq
{m}_0^2 \equiv  m^2_\gamma
+ {e^2 (3\pi^2 J_0)^{2/3} \over \pi^2}\,,~~~m_\gamma \equiv 
2e \sqrt{ J_0\over 2 m_H}\,.
\label{m0m}
\eeq
The calculation of the spectrum is a bit lengthy 
but straightforward.  The result is 
as follows: The  two (out of three) helicities  of the massive photon 
are transverse and propagate according to the massive dispersion relation
\beq
\omega^2 = {\bf k}^2 + m^2_\gamma\,, 
\label{disp1}
\eeq
where ${\bf k} $ denotes the spatial three momentum.
Furthermore, there are two modes with the dispersion relations
\beq
\omega_{\pm}^2 &=& {\bf k}^2 \left ( {m_0^2 + m^2_\gamma \over 2 m_0^2 }
\right ) + {2M^4 \over m_0^2} + {m^2_\gamma \over 2 }  \nonumber  \\
& \pm &  
\sqrt {4 {\bf k}^2 {M^4 m^2_\gamma \over m_0^4}+ \left [
{2M^4 \over m_0^2} - {m^2_\gamma \over 2 }+ {\bf k}^2
\left ( {m_0^2 - m^2_\gamma \over 2 m_0^2 }
\right ) \right]^2}\,.
\label{disp2}
\eeq
The solution with the plus subscript 
corresponds to  the scalar mode  and  its mass squared  
in this frame  is $\omega_{+}^2 ({\bf k}=0)= (4M^4/m^2_0)$, while the solution 
with the minus  subscript corresponds to the longitudinal 
component of the massive vector field, with $\omega_{-}^2 
({\bf k}=0)=m^2_\gamma$. Here it is 
useful to note that the expression for 
$M$ (\ref {M}) can be rewritten as $M=(m_H m_\gamma)^{1/2}$. 

Furthermore, in the limit when the fermions 
are non-dynamical (i.e.,  are frozen ``by hand'' or some other dynamics),
then  $m_0 \to m_\gamma$, and the  solutions reduce to the 
ones obtained in \cite {GGRR1}, however,  in  the present 
setup the difference between $m_0$ and $m_\gamma$ is greater 
than  $ m_\gamma$. Therefore, the fermion dynamics contributes by an 
additional  screening of the electrostatic interactions.

The solutions  (\ref {disp1}) and  (\ref {disp2})
are positive for arbitrary ${\bf k}$. From this, we conclude 
that the charged condensate background  
is stable w.r.t. small perturbations.
All the group  velocities obtained from  
(\ref {disp1}) and  (\ref {disp2}) 
are subluminal.

\section{Brief discussions and outlook}

The spectrum of bosons exhibits a mass gap. As a result,
contributions of the bosons into the specific heat 
of the substance at low temperatures  would be suppressed 
as  ${\rm exp} (-m_\gamma/T)$, or stronger. Since for the densities at hand  
$m_\gamma \simeq (30-90)~KeV$, (corresponding to $\sim (3 - 9)\cdot 10^8~K$)
the exponential suppression will be strong at lower temperatures.

As the star cools further, the specific heat will be dominated by the 
contributions of the relativistic electrons near the Fermi surface, 
and, hence,  would  vanish with  the temperature linearly, $\sim T$.  
This resembles the low temperature scaling in metals, 
but differs from that  in insulating solids,  or in 
the BE condensate of the neutral helium-4 atoms,  
where the low temperature specific heat is dominated by 
massless phonons, and scales as $\sim T^3$. 

If charged condensate is realized in the helium dwarf stars, 
as we have argued in this work, 
there may  be important consequences for the equation 
of state, cooling, rotation, as well as  
the asteroseismology of  the helium dwarfs.  
These and related issues will be studied elsewhere.

\vspace{0.1in}

We are grateful to Stefan Hofmann, Juan Maldacena and 
Hector Rubinstein for stimulating questions,  and to 
Nima Arkani-Hamed, Igor Klebanov and Andrew 
MacFadyen for useful discussions. GG thanks Igor Klebanov and 
Physics Department of Princeton University for hospitality, and 
the NSF and NASA for grant support (PHY-0758032,  
NNGG05GH34G). RAR acknowledges the James Arthur 
graduate fellowship support.


\begin{thebibliography}{99}

\bibitem{Shapiro}  S.L. Shapiro and S. A. Teukolsky, 
{\it ``Black Holes, White Dwarfs, and Neutron Stars''}, 
John Wiley \& Sons, (1983).

\bibitem{Ruderman} L. Mestel and M.A. Ruderman, MNRAS, {\bf 136}:27 (1967).

\bibitem{vanHorn} D.Q. Lamb and H.M. Van Horn, Ap.J. {\bf 200}:306 (1975).

\bibitem{Ichi} S. Ichimaru, Rev. Mod. Phys. {\bf 54}:1017 (1982).

\bibitem{DeWitt} G.S. Stringfellow, H.E. DeWitt, W.I. Slattery,
Phys. Lett. {\bf A 41}, 1105 (1990).
 
\bibitem{Chabrier} G. Chabrier, Ap.J. {\bf 414}:695 (1993).

\bibitem{WDs} B.M.S. Hansen and J. Liebert, Ann.Rev. 
Astron. Astrophys. {\bf 41}: 465 (2003).

\bibitem{Ashcroft} G. Chabrier, N.W. Ashcroft, and H.E. DeWitt, 
Nature, {\bf 360},48 (1992)


\bibitem{GGRR1} G.~Gabadadze and R.~A.~Rosen,
Phys Lett. {B} {\bf 658},  266 (2008.

\bibitem{GGRR2} G.~Gabadadze and R.~A.~Rosen,
  Phys.\ Lett.\  B {\bf 666}, 277 (2008)



\bibitem{Straumann} N. Straumann, {\it ``General relativity with 
Applications to Astrophysics''}, Springer-Verlag, (2004) 


\bibitem{Linde} A.~D.~Linde,
  Phys.\ Rev.\  D {\bf 14}, 3345 (1976).


\bibitem{Haber} H.~E.~Haber and H.~A.~Weldon,
  Phys.\ Rev.\  D {\bf 25}, 502 (1982).


\end{thebibliography}
\end{document}